\documentclass[conference]{IEEEtran}
\usepackage{cite}

\ifCLASSINFOpdf
   \usepackage[pdftex]{graphicx}
\else
\fi
\usepackage[cmex10]{amsmath}
\newtheorem{theorem}{Theorem}

\newtheorem{lemma}{Lemma}
\newtheorem{remark}{Remark}

\begin{document}
\title{An Achievable Rate-Distortion Region for the Multiple Descriptions Problem}

\author{
  \IEEEauthorblockN{Farhad Shirani}
  \IEEEauthorblockA{EECS Department\\University of Michigan\\ Ann Arbor,USA \\
    Email: fshirani@umich.edu} 
  \and
  \IEEEauthorblockN{S. Sandeep Pradhan}
  \IEEEauthorblockA{EECS Department\\University of Michigan\\ Ann Arbor,USA \\
    Email: pradhanv@umich.edu}
}


%


\maketitle

\begin{abstract}
A multiple-descriptions (MD) coding strategy is proposed and an inner bound to the achievable rate-distortion region is derived. The scheme utilizes linear codes. It is shown in two different MD set-ups that the linear coding scheme achieves a larger rate-distortion region than previously known random coding strategies. Furthermore, it is shown via an example that the best known random coding scheme for the set-up can be improved by including additional randomly generated codebooks. 
\end{abstract}


%
\IEEEpeerreviewmaketitle

\section{Introduction}
The multiple-descriptions (MD) source coding set-up describes a communications system consisting of a centralized encoder and several decoders. The encoder transmits data through a number of noiseless links. Each decoder is connected to the encoder via a subset of these links. The goal is for the encoder to compress an information source and transmit it to the decoders such that the source reconstruction at each decoder meets a specific fidelity criterion.
 There has been an extensive amount of effort to determine the optimal rate-distortion (RD) region for the general MD set-up, however, even in the case of two-descriptions the optimal region is not known. The best known achievable RD region for the two-descriptions set-up is due to Zhang and Berger \cite{1}. In \cite{1}, the encoder utilizes a base layer which is decoded by all receivers and a refinement layer which is decoded by individual receivers. The VKG scheme proposed in \cite{2} generalizes the base layer idea in \cite{1} to cases with more than two-descriptions. The combinatorial-message-sharing (CMS) strategy \cite{3} expands the method in \cite{2} by
 considering a combinatorial number of base layer codebooks which are decoded in subsets of receivers
. In \cite{4}, a random binning scheme was introduced which results in gains over previous known coding strategies. The method in \cite{4} is only applicable to symmetric sources. Finally, in \cite{6} the ideas in \cite{3} and \cite{4} were combined to form CMS with binning. It was shown that CMS with binning gives gains over previous coding strategies and strictly contains them. All of these coding schemes use random codes to construct codebooks; in this paper we propose using linear codes instead. 

\begin{figure}[!t]
\centering
\includegraphics[width=2.5in,height=1in]{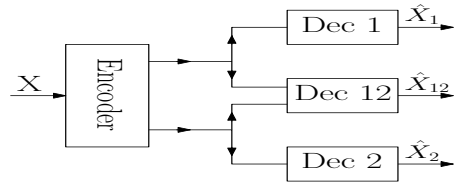}
\caption{The two-descriptions problem}
\label{fig_sim}
\end{figure}

 Using structured codes in communications problems has traditionally been of interest due to their practicality in comparison with randomly generated codes. Korner and Marton \cite{7} observed that in some set-ups, application of structured codes may also yield gains in terms of achievable rate-distortions. Specifically they show that in a particular 3-user distributed source coding problem, involving reconstruction of a sum of two BSS\rq{}s, using linear codes results in a larger achievable RD region. The phenomenon was also observed in channel coding problems. It was shown in the three user interference channel \cite{8} and the three user broadcast channel \cite{9}, that employing linear codes results in gains. Intuitively, the main idea behind all of these linear coding schemes is that because of their structure, linear codes can compress and transmit sums of binary RV\rq{}s more efficiently than random codes. Based on these observations it is expected that utilizing linear codes is also advantageous in the MD problems when more than 2 descriptions are transmitted. This turns out to be indeed the case as we illustrate in the next chapters.

The rest of the paper is organized as follows: Section II is allocated to explaining the CMS with binning scheme. In section III, we prove linear codes give gains over previous schemes in two different examples. Section IV contains a proof that the CMS with binning scheme can be improved using random codes. In section V, we provide an achievable RD region for the MD problem. Section VI concludes the paper.

\section{CMS with binning}
Here we explain the CMS with binning scheme presented in \cite{6} for the $L$-descriptions problem.

\textbf{Base Layer Construction:} For each subset $A$ of $[1:L]$, we construct $|A|$ codebooks $C_{A,i},i\in [1:|A|]$. Each codebook is generated based on the probability distribution $P_{V_{A,i}}$, independent of other codebooks. The codebook has rate $r_{A,i}$. This codebook is to be decoded if a decoder receives at least $i$ descriptions from the set $A$. For each description in $A$, the encoder bins the codebook at a different rate. Binning is done for each description independent of other descriptions. The binning rate of codebook $C_{A,i}$ for description $j$ is $\rho_{A,i,j}$. This gives bin size $r_{A,i}-\rho_{A,i,j}$. On description $j\in A$, the encoder sends the bin number of the codeword to be transmitted from $C_{A,i}$, this requires rate $\rho_{A,i,j}$. 

 \textbf{Refinement Layer Construction:} For description $j$ we construct $L-1$ refinement layer codebooks, $C_{i,j},i\in[1:L-1]$. Each codebook is generated based on $P_{U_{i,j}}$ and has rate $r_{i,j}$. The codebook is decoded if the decoder receives description $j$ along with at least $i-1$ other descriptions (i.e. the codebook is an SCEC sent by encoder $j$). The codebook is binned at rate $\rho_{i,j}$. 

\textbf{Covering Bounds:} Since the codebooks are generated independently, typicality requires mutual covering bounds for all subsets of random variables.
\\ Hence for all $\mathcal{A}=\{(A,j)|A\subset [1:L],j\in[1:|A|]\}$ and $K=\{(k,n)|k\in[1:L-1],n\in[1:L]\}$ we must have:

\begin{align*}
&H(V_\mathcal{A},U_K|X)\geq  \!\!\!  \sum_{(A,j)\in      \mathcal{A}}{  \!\!\!\!\!\!   (H(V_{{A,j}})  \!  -   \!   r_{A,j})}+     \!\!\!\!\!\!     \sum_{(k,n)\in K}{    \!\!\!\!\!\!    (H(U_{k,n})   \!    -    \!    r_{k,n})}
\end{align*}

\textbf{Packing Bounds:} For decoder $\underline{s}$, let $\mathcal{A}_{\underline{s}}$  be the indices of codebooks $C_{A,k}$ decoded at $\underline{s}$. Also let $K_{\underline{s}}$ be the indices $(k,n)$ of codebooks $C_{k,n}$ decoded at $\underline{s}$. Let $\mathcal{A}_1$ and $\mathcal{A}_2$ partition $\mathcal{A}$. Also let $K_1$ and $K_2$ partition $K$. For all such sets, we have the following packing bounds:

\begin{align*}
&H(V_{\mathcal{A}_1},U_{K_1}|V_{\mathcal{A}_2},U_{K_2})\leq \!\!\!\!\!      \sum_{(k,n)\in K_1}{\!\!\!\!(H(U_{k,n})+          \rho_{k,n}-     r_{k,n})}\\
&\qquad + \sum_{(A,j)\in \mathcal{A}_1}{  \!\!\!\!  (H(V_{{A,j}})+ (\sum_{i\in\underline{s}}\rho_{A,j,i})-r_{A,j})}).
\end{align*}
\section{Linear Coding Examples}
In this section we present two examples showing that linear codes attain points outside of previous known achievable RD regions. 
\subsection{A Three User Example}
Figure 2 depicts the three-descriptions problem. Here $X$ and $Z$ are independent BSS\rq{}s. Distortion is measured at individual decoders (i.e. decoders 1,2 and 3) using Hamming distortion. We choose the distortion functions for decoders 12,13 and 23 such that in a PtP setting they achieve optimal rate-distortion by receiving two independent quantizations of $X$ and $Z$, where the independent quantizations are done using binary symmetric test channels with cross over probability $\delta\in[0,0.5]$. To construct such a distortion function we use the method in \cite{10}. Let $P_{XZ\bar{X}\bar{Z}}$ be the joint probability distribution of the source along with two quantizations $\hat{X}=X+N_{\delta}$ and $\hat{Z}=Z+N\rq_{\delta}$ where $N_{\delta}$ and $N\rq{}_{\delta}$ are $Be(\delta)$ (i.e. with $P(N_{\delta}=1)=\delta$) and independent of all other RV\rq{}s. Then the distortion function between the source $(X,Z)$ and the reconstruction $(\hat{X},\hat{Z})$ is defined as 
\begin{align*}
&d_{XZ}((x,z),(\hat{x},\hat{z}))=-c\log{p_{XZ|\hat{X}\hat{Z}}(x,z|\hat{x},\hat{z})}+d_0(x,z).
\end{align*}
Here $d_0$ is chosen such that $d_{XZ}((x,z),(x,z))\!\!=\!\!0, \forall x,z\in \{0,1\}$. Also $c$ is an arbitrary positive constant. With this distortion function, in a PtP setting if we construct a test channel using $P_{XZ\hat{X}\hat{Z}}$,  it achieves rate-distortion at  $D=E_{P_{XZ\hat{X}\hat{Z}}}\{d_{XZ}((X,Z),(\hat{X},\hat{Z}))\}$ and $R=I(\bar{X},\bar{Z};X,Z)=2(1-h_b(\delta))$. 
\begin{figure}[!t]
\centering
\includegraphics[width=2.5in,height=2.2in]{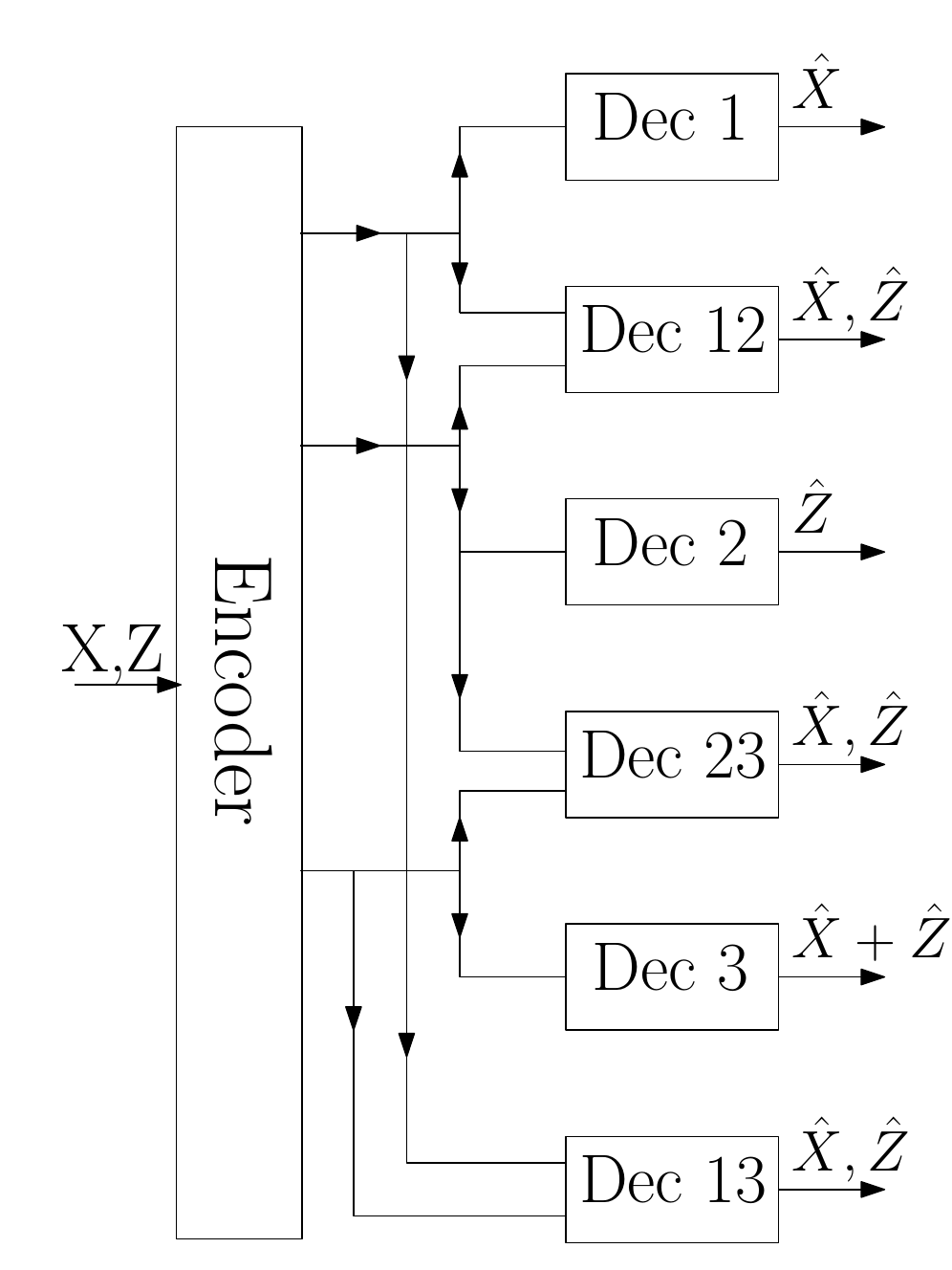}
\caption{Three-Descriptions Example}
\label{fig_sim}
\end{figure}
\theorem In the above MD problem, linear codes  can achieve the following rate-distortions: 
\begin{align*}
&R_i=1-h_b(\delta), D_1=D_2=\delta,D_3=\delta\ast\delta\\
&D_{12}=D_{13}=D_{23}=D
\end{align*}
\begin{IEEEproof} Here we propose a linear coding scheme that achieves the above rates. 
\\\textbf{Encoding:} Define $r=1-h_b(\delta)$. Let $C_{rn\times n}$ be a family of linear codes which quantize a BSS to Hamming distortion $\delta+\lambda_n$ for some $\lambda_n\to 0$. Let $G_{rn\times n}$ be the generator matrices for these linear codes. Let $U_1^n$ be the quantization of $X^n$ using $C_{rn\times n}$ (i.e. $U_1^n=argmin_{\hat{x}^n}\{d_H(x^n,\hat{x}^n)|\hat{x}^n \in C_{rn\times n}\}$). Also define $U_2^n$ to be the quantization of $Z^n$ using the same code. Note that since $C_{rn\times n}$ is a linear code, $U_1^n+U_2^n\in C_{rn\times n}$. The first description carries the index of $U_1^n$, the second description carries the index of $U_2^n$ and the third description sends the index for $U_1^n+U_2^n$  in $C_{rn\times n}$.
\\ \textbf{Decoding:} The first and second decoder get the index of $U_1^n$ and $U_2^n$ respectively and hence satisfy their distortion constraints. Decoder 3 reconstructs $U_1^n+U_2^n$, and it is easy to show that $\frac{1}{n}E(d_H(U_1^n+U_2^n,X^n+Z^n)\to \delta\ast\delta$. Decoder 12 receives $U_1^n$ and $U_2^n$ and hence satisfies its distortion requirements. Also decoders 13 and 23 can recover $U_2^n$ and $U_1^n$  by adding $U_1^n+U_2^n$ to $U_1^n$ and $U_2^n$ respectively.
\end{IEEEproof}
One may observe the main idea in the proof is that due to the linearity of the code, $U_1^n+U_2^n$ is in the codebook, hence it can be sent on the third description with the same rate as other descriptions. 
Now we prove that CMS with binning cannot achieve the above rate-distortions. We do this by assuming such a rate-distortion vector is achievable and then arriving at a contradiction. 
\theorem The rate-distortions in theorem 1 are not achievable using the CMS with binning scheme.
\begin{IEEEproof}
Figure 3 shows the codebooks present in CMS with binning for three descriptions. 
\begin{figure}[!t]
\centering
\includegraphics[width=2.5in]{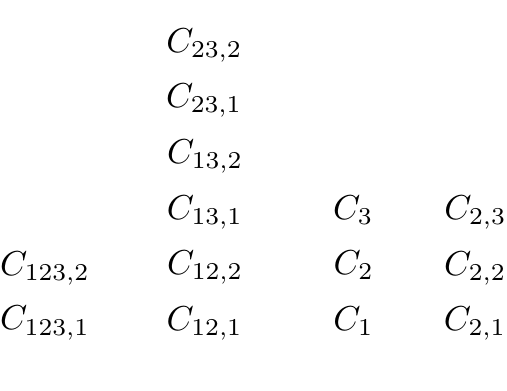}
\caption{CMS with binning for three descriptions}
\label{fig_sim}
\end{figure}
\\\textbf{Step 1:} It is straightforward to check that $\rho_{123,2,1}$, $\rho_{123,2,2}$ ,$\rho_{12,2,1}$, $\rho_{13,2,1}$, $\rho_{12,2,2}$, $\rho_{23,2,2}$, $\rho_{2,1}$ and $\rho_{2,2}$ are 0. The intuitive reason is that decoder 1 receives the first description at optimal PtP rate-distortion, hence the first description can\rq{}t carry any indices which are not used in decoder 1. Note that this does not mean the codebooks  relating to these binning rates are empty, we can only conclude that no bin indices relating to the above codebooks are sent through the corresponding descriptions.  
\\\textbf{Step 2:} The random variables decoded at decoder 1 and decoder 2 are independent of each other, because decoder 12 is operating optimally in a PtP communications point of view, hence any correlation between descriptions 1 and 2 would be redundant and would contradict optimality. To show this we investigate a more general situation in lemma 1 in the appendix. From lemma 1, even if the refinement layer is included, there is no common codebook decoded at decoders 1 and 2. Hence in our situation $C_{123,1}=C_{12,1}=\phi$. Also because of the Markov chain $C_{12,2}$ is not used in reconstructing the source in decoder 12, so it can be eliminated without any loss. $C_{123,2}$ is only sent through description 3 and is only used in decoders 13 and 23 (by the Markov chain), so it can be combined with $C_{2,3}$ and we only keep the latter. Also note that $C_{2,1}$ is not sent through any description and is only used in decoder 13, so it can be pushed into $C_{13,2}$ without any penalty (i.e. we replace $V_{13,2}$ with $(V_{13,2},U_{2,1})$). We can eliminate $C_{2,2}$ in the same manner.
\\\textbf{Step 3:} Note that decoder 23 is operating at PtP rate-distortion. Also $C_{13,2}$ is carried by description 3 through $\rho_{23,2,3}$ but not used that decoder. So by the same arguments as before $\rho_{23,2,3}=0$. Using a similar argument we deduce $\rho_{13,2,3}=0$.
\\\textbf{Step 4:} We proceed by showing that $C_{13,2}$ and $C_{23,2}$ are empty. So far it was shown that these codebooks are not transmitted through any description, however we have not shown they are empty (i.e. they are not decoded anywhere by using their correlation with other RV\rq{}s). Intuitively since these random codewords can only be decoded in a decoder through other random variables, they must not be giving any extra information about the source. To prove the redundancy of these codebooks, consider the following packing bounds for decoders 1, 23 and 13:
\begin{align}
&H(V_{13,1}U_1)\leq H(V_{13,1})+H(U_1)+R_1-r_{13,1}-r_1\\
&H(V_{13,1},V_{23,1},V_{23,2},U_{2,3},U_2,U_3)\leq H(V_{13,1})+H(V_{23,1})\nonumber\\&+H(V_{23,2})+H(U_{2,3})+H(U_2)+H(U_3)+R_2+R_3-\nonumber\\&r_{13,1}-r_{23,1}-r_{23,2}-r_{2,3}-r_2-r_3\\
&H(V_{13,2}|V_{13,1},V_{23,1},U_1,U_3,U_{2,3})\leq H(V_{13,2})-r_{13,2}
\end{align}
We add the above inequalities and subtract the mutual covering bound on all RV\rq{}s. After some simplification we get $I(X,Z;V_{13,2}|U_1,U_3,U_{2,3},V_{13,1},V_{23,1})\leq 0$. This imposes the Markov chain $V_{13,2}\leftrightarrow U_1,U_3,U_{2,3},V_{13,1},V_{23,1} \leftrightarrow X,Z$. Hence $V_{13,2}$ is not necessary for reconstructing the source at decoder 13, which means $C_{13,2}$ can be eliminated without any loss. Same argument works for eliminating $C_{23,2}$.
 \\\textbf{Step 5:} In this step we show that $\rho_{13,1,1}=\rho_{13,1,3}=\rho_{23,1,2}=\rho_{23,1,3}=0$. To see this assume $\rho_{13,1,1}> 0 $. Note that $V_{13,1}$ is decoded at decoder 3, so even if description 1 did not carry the index of the codeword in $C_{13,1,1}$, decoder 13 could decode $V_{13,1}$ using the third description and calculate the index. Hence $\rho_{13,1,1}$ could be set to 0 without any added distortion at decoder 13. This contradicts optimality at decoder 13. Now since $C_{13,1}$ and $C_{23,1}$ are not carried by any descriptions, we can use the same kind of argument as in the previous steps, by adding the packing bounds at decoders 1,3 and 13 and subtracting the mutual covering bound on all variables, we get that $r_{12,1}=0$. 
\\\textbf{Step 6:} We are left with four codebooks, $C_1,C_2,C_3$ and $C_{2,3}$. Note that since decoder 1 is only decoding $C_1$ we must have $\rho_1=r_1=R_1$. This is deduced from the packing bound in decoder 1:
\begin{equation*}
H(U_1)\leq H(U_1)+\rho_1-r_1\to r_1\leq \rho_1
\end{equation*}
But $\rho_1\leq r_1$ so they are equal. The same argument gives $\rho_2=r_2=R_2$. Also $\rho_3=r_3$ and $R_3=r_3+\rho_{2,3}$. We have the following packing bound at decoder 13:
\begin {align}
&H(U_1,U_3,U_{2,3})\leq H(U_1)+H(U_3)+H(U_{2,3})+R_1+R_3\nonumber\\
&\qquad\qquad\qquad\qquad\qquad\qquad\qquad-r_1-r_3-r_{2,3}\nonumber\\
&\to r_{2,3}-\rho_{2,3}\leq I(U_3;U_{2,3})+I(U_1;U_3,U_{2,3}).
\label{end}
\end{align}
Where we have used $R_1+R_3=I(U_1,U_3,U_{2,3};X,Z)$ from optimality at decoder 13. Adding inequality \eqref{end} with the mutual covering bound on all variables we get:
\begin{align}
&R_3\geq I(U_3,U_{2,3};XZU_1U_2)-I(U_1;U_3,U_{2,3})\nonumber\\
&\to R_3\geq H(XZU_2|U_1)-H(XZU_2|U_1U_3U_{2,3})\nonumber\\
&\to R_3\geq H(U_2)-H(U_2|U_1U_3U_{2,3}XZ) 
\label{markov}
\end{align}
Where in the last step we have used $H(XZ|U_1U_2)=H(XZ|U_1U_3U_{2,3})$. Note that $I(U_2;X)=1-h_b(\delta)$, so the RHS in the last equality is greater than or equal $1-h_b(\delta)$. Equality requires that $U_3,U_{2,3}\leftrightarrow X,Z,U_1\leftrightarrow U_2$. By the same arguments we get $U_3,U_{2,3}\leftrightarrow X,Z,U_2\leftrightarrow U_1$. Now we use the second lemma in the appendix. Let $A=(U_3,U_{2,3})$, $B=(X,Z)$, $C=U_1$ and $D=U_2$ in the lemma. The conditions of the lemma are indeed true, because $U_1\leftrightarrow X,Z \leftrightarrow U_2$, hence we can\rq{}t have functions $f_{x,z}$ and $g_{x,z}$ which are equal with probability 1. Using the lemma, the following Markov chain holds $U_3,U_{2,3}\leftrightarrow X,Z\leftrightarrow U_1,U_2$.
Recall in \eqref{markov} we used the mutual covering bound on all variables and since we proved all of the inequalities used in that part need to be equalities, the mutual covering bound is tight. Also from optimality of decoder 23 we get that the covering bound on $U_2,U_3$ and $U_{2,3}$ is tight. Subtracting these two equalities we get $H(U_1|U_2U_3U_{2,3}XZ)=H(U_1|X)$, so we must have $U_1\leftrightarrow X \leftrightarrow U_2,U_3,U_{2,3},Z$. Also from the definition of $d_{13}$ we must have $Z\leftrightarrow U_1,U_3,U_{2,3}\leftrightarrow X$, so $I(Z;X|U_1U_3U_{2,3})=0$. Then we have:
\begin{equation*}
I(U_1;X)=I(U_1,U_3,U_{2,3};X)=I(Z,U_1,U_3,U_{2,3};X)
\end{equation*}
Where the second equality holds since both sides are equal to $1-h_b(\delta)$ since they give reconstructions of $X$ at decoders 1 and 13. So we get $X\leftrightarrow U_1\leftrightarrow Z,U_3,U_{2,3}$. By lemma 2, $X,U_1\perp Z,U_3,U_{2,3}$ holds (take $A=(U_3,U_{2,3}),B=\phi,C=X,D=U_1$).
In this case, decoder 3 can reconstruct both $X$ and $Z$ with Hamming distortion $\delta$ this contradicts $R_3=1-h_b(\delta)$. To get the reconstructions at decoder 3, let $g(U_1,U_3,U_{2,3})$ be the reconstruction of $Z$ at decoder 13. We have:
\begin{align*}
&\sum_{z,u_1,u_3,u_{2,3}}p(z,u_1,u_3,u_{2,3})d_H(g(u_1,u_3,u_{2,3}),z)\leq \delta\to\\
&\to \sum_{u_1} p(u_1)\sum_{z,u_3,u_{2,3}}p(z,u_3,u_{2,3})d_H(g(u_1,u_3,u_{2,3}),z)\leq \delta\\
\end{align*}
So there is at least one $u_1\in \mathcal{U}_1$ such that $\sum_{z,u_3,u_{2,3}}p(z,u_3,u_{2,3})d_H(g(u_1,u_3,u_{2,3},z)\leq \delta$. Let $g_{u_1}(U_3,U_{2,3})=g(u_1,U_3,U_{2,3})$ be the reconstruction of Z using $U_3$ and $U_{2,3}$. By the same argument we can find a reconstruction of X.
\end{IEEEproof}
\subsection{A Four-Descriptions Example}
So far we proved linear codes outperform previous random coding schemes in the three-descriptions problem. The gains are only presenting themselves due to the fact that linear codes can compress sums of binary RV\rq{}s more efficiently, these are the same gains as the ones in other three-terminal communications problems. Now we proceed to explain our second example. The example involves a four-descriptions problem. We believe the gains in this example point out to a new phenomenon which arises when using linear codes. The set-up is depicted in figure 4. Here $X$ and $Z$ are BSS\rq{}s which are related to each other through a BSC$(\delta)$ (i.e. $X=Z+N_{\delta}$ where $N_\delta$ is $Be(\delta)$ and independent of $X$ and $Z$). We are interested in the operating point where decoder 1 reconstructs X with Hamming distortion $\delta$, decoder 4 reconstructs $Z$ with the same distortion, the rest of the reconstructions are lossless as shown in the figure. 
\begin{figure}[!t]
\centering
\includegraphics[width=2.5in]{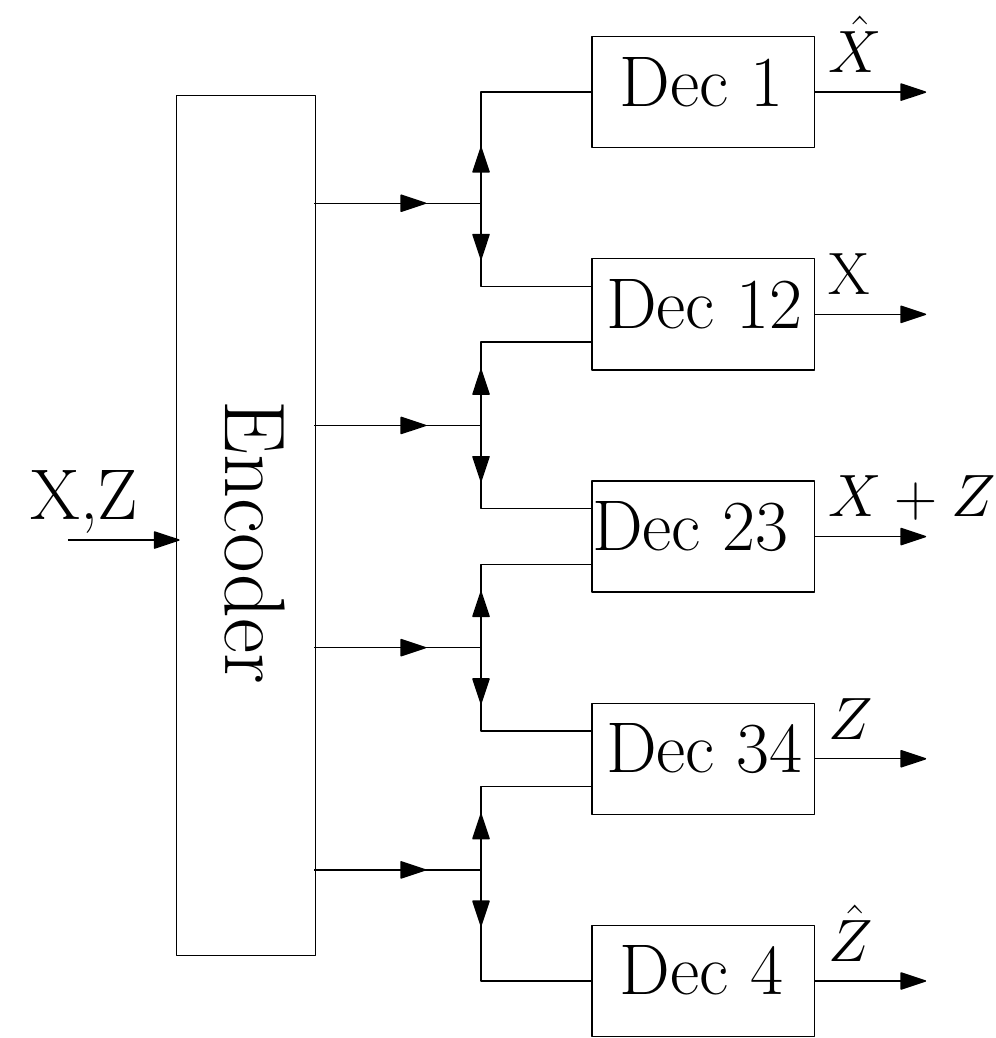}
\caption{Four-descriptions example}
\label{fig4}
\end{figure}
\theorem For the above distortions, linear codes achieve the following rates:
\begin{equation*}
R_1=R_4=1-h_b(\delta), R_2=R_3=h_b(\delta)
\end{equation*}
\begin{IEEEproof}
We proceed by presenting a linear coding scheme which achieves the above rates and distortions. Let $C_{rn\times n}$ and $G_{rn\times n}$ be defined as in the previous example. The only difference is here we assume that $C_{rn\times n}$ is both a good channel code for a BSC$(\delta)$ and a good source code for quantizing a BSS to Hamming distortion $(\delta+\lambda_n)$ where $\lambda_n\to 0$. The existence of such codes can be proved using a simple Shannon type argument. 
\\\textbf{Encoding:} The encooder quantizes $X^n$ using $C_{rn\times n}$ to $u^k$ and sends the index on description 1. It also quantizes $Z^n$ using the same code to $v^k$ and sends the index on description 4. The quantization noise at decoder 1, $X^n+u^kG_{rn\times n}$, is sent on description 2, also the quantization noise at decoder 4 is sent on description 3. Clearly the rates are as stated in the theorem.
\\\textbf{Decoding:} Decoder 1 and 4 are receiving their desired quantizations. Decoder 12 adds the quantization $u^kG_{rn\times n}$ of $X^n$ to its quantization noise to recover $X^n$ noiselessly. Decoder 34 recovers $Z^n$ in the same manner. Decoder 23 gets the two quantization noises. It then adds the two to get $(X+Z)^n+(u+v)^kG_{rn\times n}$, treating $X+Z$ as noise it can decode $(u+v)^k$ since the code is a good channel code for $BSC(\delta)$ and hence it can then reconstruct $(X+Z)^n$ noiselessly (the careful reader might notice with rate exactly $1-h_b(\delta)$ the code can only be a good channel code for channels with crossover probability strictly less than $\delta$, while there is a suitable fix to this issue, one can totally bypass it by assuming $X+Z$ is $Be(\delta-\lambda)$ for some small positive $\lambda$).
\end{IEEEproof}
Note the linearity of the codebook, along with it being a good channel code and a good source code are crucial for achieving this RD vector. Now we prove that CMS with binning does not achieve the rates and distortions in the previous theorem.
\theorem CMS with binning does not achieve the RD vector in theorem 3. 
\begin{IEEEproof}
Again we prove the theorem by assuming the RD vector is achievable and arriving at a contradiction. The CMS scheme uses 39 codebooks for the four-descriptions problem, however since in the special case which we are considering a large number of decoders are not present, the codebooks can be eliminated in a straightforward fashion
\\\textbf{Step 1:} Any codebook which is not decoded at decoders 1,4,12,23 and 34 is redundant. For example there are no decoders receiving more than two descriptions, so any  codebook which is decoded only when three or more descriptions are available is redundant. 
\\\textbf{Step 2:}   By the same kind of argument as in lemma 1, we can show there is nothing common decoded at decoders 12 and 34. also by the same arguments as in step 2 of the last part, $C_{2,1}$ and $C_{2,4}$ can be eliminated. 
\\\textbf{Step 3:} Note since decoders 2 and 3 are not present, $C_2$ and $C_3$ are the same as $C_{2,2}$ and $C_{2,3}$, so we only keep the two latter codebooks. 
\\\textbf{Step 4:} By the same arguments as in step 3 of the last proof $C_{123,2}$ and $C_{234,2}$ are not sent through any description. By the same type of calculations as in step 4 of the last part, they can be eliminated.
\\\textbf{Step 5:}  The 8 remaining codebooks are $C_1$, $C_{2,2}$, $C_{2,3}$, $C_{4}$, $C_{12,1}$, $C_{12,2}$, $C_{34,1}$ and $C_{34,2}$. In this step we eliminate the last four codebooks. We have the following packing bounds in decoders 1 and 12:
\begin{align*}
&H(V_{12,1},U_1)\leq H(V_{12,1})+H(U_1)+R_1-r_{12,1}-r_1\\
&H(V_{12,2},U_{2,2}|U_1,V_{12,1})\leq H(V_{12,2})+H(U_{2,2})+\\&R_2-\rho_{12,1,2}-\rho_{23,2,2}-r_{2,2}-r_{12,2}\\
\end{align*}
We add these bounds and subtract the mutual covering bound on $U_1,V_{12,1},V_{12,2}$ and $U_{2,2}$. After some simplification and using the fact that decoder 12 can reconstruct $X$, we get that $\rho_{12,1,2}=\rho_{23,2,2}=0$. Also $\rho_{12,2,2}=0$ to see this, consider decoders 23 and 34, if we consider them as a joint decoder, they are performing at PtP rate-distortion, but are not decoding $C_{12,2}$, so description 2 can\rq{}t carry this codebook, hence the codebook is not sent through any descriptions and using the same arguments as in the previous proof it can be eliminated. Now after eliminating $C_{12,2}$ it is simple to eliminate $C_{12,1}$. Consider the following packing bounds at decoders 1 and 23:
\begin{align*}
&H(V_{12,1},U_1)\leq H(V_{12,1})+H(U_1)+R_1-r_{12,1}-r_1\\
&H(U_{2,2},U_{2,3},V_{12,1},V_{34,1},V_{23,2})\leq H(U_{2,2})+H(U_{2,3})+\\&H(V_{12,1})+H(V_{34,1})+H(U_{23,2})+R_2+R_3-\\&r_{2,2}-r_{2,3}-r_{12,1}-r_{34,1}-r_{23,2} 
\end{align*}
Add the two packing bounds and subtract the mutual covering bound on $U_1,U_{2,2},U_{2,3},V_{12,1},V_{12,2},V_{23,2}$ and $V_{34,1}$ to get $r_{12,1}=0$ (Here we use the fact that having all the variables decoded at decoders 23 and 12 we are able to reconstruct $(X,Z)$ so $I(XZ;U_1,U_{2,2},U_{2,3},V_{12,1},V_{34,1})=1+h_b(\delta)$). Also using the same bounds in decoders 4 and 23, we can show $r_{34,1}=0$. 
\\\textbf{Step 6:} By an argument like the one in lemma 1 we can show that considering decoders 12 and 34 we must have $U_{2,2}\leftrightarrow X,Z \leftrightarrow U_{2,3}$, also at decoder 12 we must have $U_{2,2}\leftrightarrow X \leftrightarrow Z$ and at decoder 34 we get $U_{3,2}\leftrightarrow Z \leftrightarrow X$. Taking $A=U_{2,2},B=X,C=Z,D=U_{2,3}$ in lemma 3, the long Markov chain $U_{2,2}\leftrightarrow X \leftrightarrow Z \leftrightarrow  U_{2,3}$ must hold. We get an inner bound for $R_2+R_3$ at decoder 23:
\begin{equation*}
R_2+R_3\geq min(I(U_{2,2},U_{2,3};X,Z))
\end{equation*}
Where the minimum is taken over all $P_{U_{2,2},U_{2,3}|X,Z}$ for which the long Markov chain is satisfied and $(U_{2,2},U_{2,3})$ give a lossless reconstruction of $X+Z$. This resembles the distributed source coding problem in \cite{7}. By the converse in that paper $R_2+R_3>2h_b(\delta)$. So the RD vector can\rq{}t be achieved using random codes. 
\end{IEEEproof} 
\section{Random Coding Improvements}
In this section we illustrate that CMS with binning can be improved by including additional randomly generated codebooks. For example the scheme does not include a codebook which is decoded when either description 1 or both descriptions 2 and 3 are received. In the situation depicted in figure 5, the addition of such a codebook results in a larger achievable RD region. Here decoders 1, 23 and 123 have Hamming distortion constraints. The distortion constraint in decoder 2 will be defined later. If decoder 2 is omitted, the example would become equivalent to the two descriptions problem discussed in \cite{1} by combining descriptions 2 and 3 into one description. In that paper, it was proved that the presence of a codebook decoded at all decoders would result in gains in achievable RD. Let $\mathcal{P}=\{P_{X_0,X_1,X_2,X}\}$ be the set of optimizing distributions in the Zhang-Berger RD region in \cite{1}, for a given $R$, $D$ and $D_0$. $X_0$ is the RV relating to the common codebook in that problem. Define $P=argmin(I(X;X_0))$, where the minimum is taken over all $P_{X_0,X_1,X_2,X}\in \mathcal{P}$.
Let $\hat{X}_0$ be an RV such that: 
\begin{equation*}
P(\hat{X}_0=\hat{x}_0|X_0=x_0)= \left\{
  \begin{array}{l l}
    p & \quad {\hat{x}_0=x_0}\\
    \frac{1-p}{|\mathcal{X}_0|-1} & \quad \text{O.W.}
  \end{array} \right.\]
Define $P_{\hat{X}_0,X}$ based on $P_{\hat{X}_0,X_0}$ and the Markov chain $\hat{X}_0\leftrightarrow X_0\leftrightarrow X$. The distortion function at decoder 2 is defined such that $P_{\hat{X}_0,X}$ is an optimizing distribution for the distortion function in a PtP setting. We are interested in achieving the following RD vector:
\begin{figure}[!t]
\centering
\includegraphics{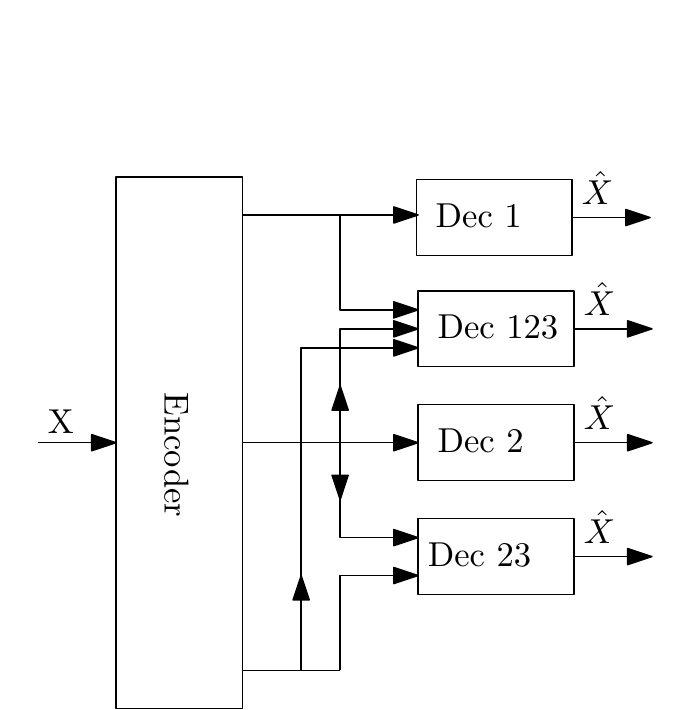}
\caption{Three descriptions exmaple for CMS with binning}
\label{fig4}
\end{figure}
\begin{align}
&R_1=R, R_2=I(\hat{X}_0;X), R_3=R-I(\hat{X}_0;X)\\
&D_1=D_{23}=D, D_2=d, D_{123}=D_0.  \nonumber
\label{RD}
\end{align}
Where $d=E_{P_{\hat{X}_0,X}}\big(d_{12}(\hat{X}_0,X)\big)$.
\theorem The above RD vector is achievable using the CMS with binning scheme with the additional codebook included.
\begin{IEEEproof}
 Define $C\rq{}$ as the codebook decoded only at decoders $1$, $23$ and $123$. Let the underlying random variable for $C\rq{}$ be $U\rq{}$. Define $N_p=X_0+\hat{X}_0$ where the addition is modulo $|\mathcal{X}_0|$. The above RD vector can be achieved by taking $U_1=X_1$, $U_2=X_0+N_p$, $U\rq{}=(X_0,X_2)$, where the distribution on $X_1$, $X_2$ and $X_0$ is $P$. 
\end{IEEEproof}
\theorem The RD vector in (1) is not achievable using CMS with binning.
\begin{IEEEproof}
Due to space limitations we only provide a summary of the proof. The common codebooks between decodes 1 and 23 are $C_{12,1}$, $C_{13,1}$ and $C_{123,1}$. Furthermore since decoder 3 is not present, $C_{123,1}$ is the same as $C_{12,1}$ and can be eliminated. We conclude that the common RV must either be sent through $C_{12,1}$ or $C_{13,1}$. Since decoder 2 is operating at optimal PtP rate-distortion, descriptions 2 and 3 can\rq{}t time-share in transmitting $X_0$ on $C_{12,1}$ and $C_{13,1}$. Let $R_0$ be the rate of the common component then by construction $R_0\geq I(X,X_0)$. So $X_0$ can\rq{}t be transmitted on either of $C_{12,1}$ and  $C_{13,1}$, which is a contradiction. 
\end{IEEEproof}
\section{Linear Coding Achievable Region}
In this section we provide an inner bound to the achievable RD region using linear codes. 
\theorem RD vectors satisfying the following bounds are achievable using linear codes. 
 Let $V=\{V_{(A,k)},(A,k)\in\mathcal{A}\}$ and $U=\{U_{(k,n)},(k,n)\in K\}$
\begin{align*}
&H(V_\mathcal{A},U_K|X)\geq  \!\!\!  \sum_{(A,j)\in      \mathcal{A}}{  \!\!\!\!\!\!   (q-\rho\rq{}_{A,j}  \!  -   \!   r_{A,j})}+     \!\!\!\!\!\!     \sum_{(k,n)\in K}{    \!\!\!\!\!\!    (q-\rho\rq{}_{k,n}   \!    -    \!    r_{k,n})}\\
\\&H(V_{\mathcal{A}_1},U_{K_1}|V_{\mathcal{A}_2},U_{K_2})\leq \!\!\!\!\!      \sum_{(k,n)\in K_1}{\!\!\!\!(q-\rho\rq{}_{k,n} +  \rho_{k,n}-     r_{k,n})}\\
&\qquad + \sum_{(A,j)\in \mathcal{A}_1}{  \!\!\!\!  (q-\rho\rq{}_{A,j} + (\sum_{i\in\underline{s}}\rho_{A,j,i})-r_{A,j})})\\
&\rho\rq{}_{A,j}\leq q-H(V_{A,j})\\
&\rho\rq{}_{k,n}\leq q-H_{U_{k,n}}.
\end{align*}
Here $q$ is the maximum of the cardinality of all RV\rq{}s involved in the optimization. Also $\underline{s}$, $\mathcal{A}_{\underline{s}}$, $K_{\underline{s}}$, $\mathcal{A}_1$, $\mathcal{A}_2$, $K_1$ and $K_2$ are defined in previous sections. 

Furthermore if the encoder wants to transmit the sum of two random variables $Y,Z\in \{U_{A,j},V_{k,n}\}$, the following covering bound must hold:  
\begin{align*}
\max\{r_y,r_z\}\geq q-H(Y+Z)
\end{align*}
If decoder $\underline{s}$ is to reconstruct $Y+Z$, then we have three cases:\\
Case 1: Decoder $\underline{s}$ reconstructs both $Y$ and $Z$. In this case, in the packing bound corresponding to this decoder, $\rho_{Y}$ is replaced with $\rho_{Y}+t\rho_{Y+Z}$ and $\rho_{Z}$ is replaced with $\rho_{Z}+(1-t)\rho_{Y+Z}$, where $t\in [0,1]$ and $\rho_{Y+Z}$ is the rate with which the codebook for $Y+Z$ is binned. 
\\Case 2: The decoder only reconstructs Y (or Z), in which case reconstructing $Y+Z$ is the same as reconstructing $(Y,Z)$. The packing bounds are written as if $Z$ was sent to the decoder with binning rate $\rho_{Y+Z}$. 
\\Case 3: The decoder does not reconstruct $Y$ or $Z$. In this case the packing bound is deduced by replacing $U_\mathcal{A}$ with $(Y+Z,U_\mathcal{A})$. 
\remark If $Y+Z$ is taken to be trivial, the above bound reduces to the CMS with binning achievable region.
\remark  $q$, $\rho\rq{}_{A,j}$ and $\rho\rq{}_{k,n}$ are eliminated after the Fourier-Motzkin elimination and do not play a role in determining the achievable region.
\remark The above rate region can be improved upon by adding the extra codebooks mentioned in the last section, and also by allowing reconstruction of multi-variate summations of the random variables.

The proof of the theorem follows from the proof of CMS with binning and simple linear coding arguments.

\section{Conclusion}
A new coding scheme for the general MD problem was proposed. It was shown that the scheme outperforms previous known random coding schemes. An example was given illustrating that previous random coding schemes can also be improved by including additional randomly generated codebooks.

\appendix
Consider the set-up in figure 1, assume $R_1+R_2=RD_{d_{12}}(D_{12})$ where $d_{12}$ is the distortion function at decoder 12, $RD_{d_{12}}$ is the PtP rate-distortion function and $D_{12}$ is distortion at that decoder. Also assume $R_i=RD_{d_i}(D_i)$. In this situation we have the following lemma:
\lemma   In CMS with binning, with the redundant refinement layer included, at the above rate-distortion vector, we must have $U_1\perp U_2$ and $C_{12,1}=\phi$. Furthermore $(U_{2,1},U_{2,2},V_{12,2})\leftrightarrow U_1,U_2 \leftrightarrow X$.
\begin{IEEEproof} Note, in this situation $C_{2,1},C_{2,2}$ and $C_{12,2}$ are only decoded at decoder 12, we define a random vector $U_0=(U_{2,1},U_{2,2},V_{12,2})$, this is the random variable which is only decoded at decoder 12. We have the following packing bounds:
\begin{align}
&H(V_{12,1},U_1)\leq H(V_{12,1}) \! + \! H(U_1) \! + \! \rho_{12,1,1} \! + \! \rho_1 \! - \! r_{12,1} \! - \! r_1\\
&H(V_{12,1},U_2)\leq H(V_{12,1}) \! + \! H(U_2) \! + \! \rho_{12,1,2} \! + \! \rho_2 \! - \! r_{12,1} \! - \! r_2\\
&H(U_0|V_{12,1},U_1,U_2)\leq H(U_0) \! + \! \rho_{0,1 \! }+ \! \rho_{0,2} \! - \! r_{0}
\end{align}

Also the covering bound:
\begin{align}
&H(U_0,U_1,U_2,V_{12,1}|X)\geq H(U_0)+H(U_1)+H(U_2)\nonumber\\
&\qquad \qquad \qquad+H(V_{12,1})-r_0-r_1-r_2-r_{12,1}
\end{align}
Now we add inequalities (1-3) and subtract the last inequality, we get:
\begin{equation*}
I(U_1;U_2|V_{12,1})+r_{12,1}\leq 0
\end{equation*}
Since both elements in the LHS are positive, both must be 0. This means $C_{12,1}=\phi$, hence $V_{12,1}$ is constant and $U_1\perp U_2$. Note that in this case some calculation reveals:
\begin{align*}
&R_1+R_2= I(U_0,U_1,U_2;X)\\
&= I(U_1;X)\!+\!I(U_2;X)\!+\!I(U_1;U_2|X)\!+\!I(U_0;X|U_1,U_2)\\
&=R_1+R_2+\!I(U_1;U_2|X)\!+\!I(U_0;X|U_1,U_2)
\end{align*}
So we must have $I(U_0;X|U_1,U_2)=0$, which gives the desired Markov chain.
\end{IEEEproof}
\lemma Let A,B,C and D be RV\rq{}s such that $A\leftrightarrow B,C\leftrightarrow D$ and $A\leftrightarrow B,D\leftrightarrow C$, and also assume there is no $b\in \mathcal{B}$ for which given $B=b$ there are non-constant functions $f_b(C)$ and $g_b(D)$ with $f_b(C)=g_b(D)$ with probability 1. Then $A\leftrightarrow B\leftrightarrow C,D$.
\begin{IEEEproof} This lemma is a generalization of the one in \cite{11}. We need to show that $p(A=a|B=b,C=c,D=d)=p(A=a|B=b,C=c\rq{},D=d\rq{})$ for any $a,b,c,c\rq{},d,d\rq{}$. Note since functions $f_b$ and $g_b$ do not exist, it is straightforward to show that there is a finite sequence of pairs $(c_i,d_i)$ such that $(c_1,d_1)=(c,d)$ and $(c_n,d_n)=(c\rq{},d\rq{})$ with the property that either $c_{i}=c_{i+1}$ or $d_i=d_{i+1}$ and that $p(B=b,C=c_i,D=d_i) \neq 0$. Then from the first Markov chain if $d_i=d_{i+1}$, we have $p(A=a|B=b,C=c_i,D=d_i)=p(A=a|B=b,C=c_{i+1},D=d_{i+1})$, also if $c_i=c_{i+1}$ the second Markov chain gives this result. So  $p(A=a|B=b,C=c_i,D=d_i)$ is constant on all of the sequence particularly $p(A=a|B=b,C=c,D=d)=p(A=a|B=b,C=c\rq{},D=d\rq{})$. 
\end{IEEEproof}
\lemma For random variables A,B,C,D, the three short Markov chains $A\leftrightarrow B,C\leftrightarrow D$, $A\leftrightarrow B \leftrightarrow C$ and $B\leftrightarrow C \leftrightarrow D$ are equivalent to the long Markov chain $A\leftrightarrow B \leftrightarrow C \leftrightarrow D$. 
\begin{IEEEproof}
We only need to show that $A\leftrightarrow B \leftrightarrow D$, the rest of the implications of the long Markov chain are either direct results of the three short Markov chains or follow by symmetry. For arbitrary $a,b,d$ we have:
\begin{align*}
&P(D=d|B=b,A=a)\\
&\!=\!\sum_{c\in{\mathcal{C}}}\!\!P(C=c|B=b,A=a)P(D=d|A=a,B=b,C=c)\\
&\!=\sum_{c\in{\mathcal{C}}}P(C=c|B=b)P(D=d|B=b,C=c)\\
&\!=P(D=d|B=b)
\end{align*}
\end{IEEEproof}




%

\end{document}